\documentclass[12pt]{article}
\usepackage{graphicx}
\usepackage{color}
\newcommand{\beq}{\begin{equation}}
\newcommand{\eeq}{\end{equation}}
\newcommand{\bea}{\begin{eqnarray}}
\newcommand{\eea}{\end{eqnarray}}

\newcommand{\gsim}{\lower.7ex\hbox{$
\;\stackrel{\textstyle>}{\sim}\;$}}
\newcommand{\lsim}{\lower.7ex\hbox{$
\;\stackrel{\textstyle<}{\sim}\;$}}

\newcommand{\eod}{\end{document}}

\definecolor{verm}{rgb}{0.8,0.1,0.0}

\include{chref}

\begin{document}
\thispagestyle{empty}
\vspace*{-22mm}

\begin{flushright}

UND-HEP-2021-BIG\hspace*{.08em}02\\
Version 2.00
 \\
%\today
December 31, 2021
%hep-ph/0703132\\
\end{flushright}

\vspace*{0.2mm}

\begin{center}

{\Large {\bf Probing violation of {\bf CP} \& {\bf T} invariance in the transitions of $\tau$ leptons
 \footnote{Contributed talk for the TAU2021 WS at the University of Indiana (U.S.A.)}}}

\vspace*{3.mm}

{\bf I.I.~Bigi }\\
\vspace{1mm}
{\sl Department of Physics, University of Notre Dame du Lac, Notre Dame, IN 46556, USA} 

\vspace*{-.8mm}

{\sl email address: ibigi@nd.edu} 
.
\vspace*{5mm} 

My motto: {\em "Crafted with Care"}
\footnote{I have `stolen' it from a local pizza restaurant in Pearland (Texas)'.}

\vspace*{5mm}

{\bf Abstract}  

\end{center}

\noindent
The SM predicts zero value for {\bf CP} \& {\bf T} violation in the decays of $\tau$ leptons --  
except $\tau^- \to \nu K_S [\pi^-/\pi^-\pi^0 ...]$. Our community could establish impact of 
New Dynamics (ND) in a {\bf CP} asymmetry in a semi-hadronic $\tau$  transition and later in a second one. 
To be `practical', I suggest to our experimental colleagues to probe $\tau^- \to \nu \bar K^0 [\pi^-/\pi^-\pi^0/\pi^-\eta]$
and $\tau^- \to \nu K^-[\pi^0/\pi^+\pi^-/\eta]$. 
While I had mostly given comments  about {\bf CP} violation in $\tau$ transitions, some theorists have shown  
how ND models could produce {\bf CP} asymmetries; I will discuss these ones, although sometimes I am not 
a true theorist, but as a phenomenalist here. 
%I am acting as a phenomenalist here. 
I can hardly `dream'  that {\bf CP} asymmetry 
in $\tau$ decays could be connected with our huge asymmetry in `our' matter vs. anti-matter. 

%%%%%%%%%%%%%%%%%%%%% 
\section{Introduction to {\bf CP} violation in $\Delta L \neq 0$ transitions}
\label{INTRO}
%%%%%%%%%%%%%%%%%%%%%%

Old people like me often start with a `history'. In this case I first give a reference to my July 2021 book with two co-authors
\cite{BRPJuly2021}. It is dedicated to Lev Okun, who said in his 1963 Russian book  `we' have to continue probing {\bf CP} violation 
in $K_L \to \pi\pi$,
 
Furthermore the TAU2021 WS is dedicated to Prof. Simon Eidelman (1948 - 2021) -- which him I had discussions for many years, and I 
had learnt \& enjoyed  -- and Prof. Olga Igonkina (1973 - 2019), when I had listened to her excellent talks at Capri (Italy) and CERN.

My 2021 book \cite{BRPJuly2021}  discusses {\bf CP} asymmetries (plus strong {\bf CP} violation, Axions, rare decays, Dark Matter, modern cosmology ...) 
about the transitions of hadrons \& leptons in general. At this TAU2021 Workshop  we have seen nice introduction talks about $\tau$  dynamics from 
Toni Pich \& Bill Marciano including its `history' (\& future challenges) as one can see on their slides \cite{TONIBILL} . 
Now one can read Toni's contribution to the Proceedings \cite{TONICON}.     
Furthermore, one can read several good talks about neutrino oscillations and EDMs. Obviously  I focus on {\bf CP}  violation in leptonic transitions.  
`My judgment' said:  With neutrino oscillations being established between three neutrinos `we' will find {\bf CP} asymmetries there;  
I would give the {\em golden} medal. Next I will give the {\em silver} medals for the EDMs for electrons, muons and tau, 
although these competitions are `close'. `We' have candidates for the {\em bronze} medal; of course, I am biased, namely to probe 
{\bf CP} asymmetries in $\tau$ decays. I discuss next.

%%%%%%%%%%%%%%%%%%%%% 
\section{{\bf CP} asymmetries in $\tau$ decays}
\label{TAUCPV}
%%%%%%%%%%%%%%%%%%%%%%

Mostly I am thinking about "leptogenesis" driving "barygenesis"; however, {\bf CP} violation in the decays of $\tau$ leptons can hardly `dream' about 
"leptogenesis" leading to "matter" $\gg$ "anti-matter" in `our' Universe. 
On the other hand, it is a wonderful hunting area for the impact of ND on {\bf CP} \& {\bf T} asymmetries (in the world of theorists). 
Their hadronic final states are enough `complex' to show impact of non-perturbative QCD  at least. It is not surprised 
that data of $\tau^- \to \nu K_S \pi^- $  has shown signs of $\bf CP$ asymmetry. The SM predicts 
$A_{\bf CP}(\tau^- \to \nu K_S \pi^- )_{\rm SM}  =  - \; (0.36 \pm 0.01) \% $ due indirect {\bf CP} violation in $K_S \to \pi^+\pi^-$. 
PDG2020 has accepted  BaBar2012 analysis leading to  $A_{\bf CP}(\tau^- \to \nu K_S \pi^- )_{\rm PDG2020}  =  + \; (0.36 \pm 0.25) \%  $: 
it is consistent  with the SM prediction -- or a sign of impact from ND.   

%%%%%%%%%%%%%%%%%%%%%%
\subsection{The theoretical `Landscape'}
\label{LANDSCAPE}
%%%%%%%%%%%%%%%%%%%%%%

`We' had listened a nice talk by Noel at the TAU2021 Workshop; one can read it in the Proceedungs \cite{Noeltalk}  
%on 9/29/2021 at 9:25 am at the TAU2021 Workshop \cite{Noeltalk} 
\footnote{My main point is the title: "On the scalar $\pi K$ form factor beyond the elastic region" both Noel's talk \& his article \cite{DETTEN}. 
First he discussed the tools for analyzing data. The truly new stuff (in my view) is the end, namely to probe {\bf CP} asymmetries in $\tau$ decays.}.  
It is based on his article \cite{DETTEN}. His talk was clearer (in my view) than in his article about finding {\bf CP} violation there, 
but also a sign of ND.  
 
On my 2009 book \cite{BSBook} one can find a reference [341] on its page 406, namely non-minimal Higgs models. Those could 
reach the $10^{-3}$ level. SUSY models might go beyond that, in particular with broken {\em R} parity \cite{DELEPINE}.

There is a list of articles by my colleague Morozumi \& his team to discuss {\bf CP} asymmetries in $\tau^- \to \nu K [\pi/\eta/\eta^{\prime}]$. 
It is not just an idea about {\bf CP} asymmetries, but to analyze some models like two Higgs doublet models. 
These ones would produce impact on forward vs. backward transitions; furthermore these models could also produce 
{\bf CP} asymmetries there  \cite{MOROZUMI}. It is very interesting in principle; however, it suggested {\bf CP} asymmetries on the level 
of ${\cal O}(10^{-6})$. Still one can learn lessons from failures of impact of ND.  
Anyway, none of these had been participants at the TAU2021 WS.

%%%%%%%%%%%%%%%%%%%%%%
\subsection{Probing {\bf CP} \& {\bf T} asymmetries in general} 
\label{PROBCPVGENERAL}
%%%%%%%%%%%%%%%%%%%%

Just above I had listed three candidates for ND to produce {\bf CP} (\& {\bf T}) asymmetries. However, I do not act as a true theorist here, but as a 
phenomenalist about {\bf CP} violation in $\tau$ decays. One example is the "Sect.5 Dynamics of $\tau$ leptons" in my Ref.\cite{TIM} with a small comment 
about future regional asymmetries from Belle II (and a possible Super Tau-Charm Factory); I will come back to that. 

I have realized that my 2021 book has giving less than one page for 
{\bf CP} asymmetry in $\tau$ decays, while around three pages in my 2009 book, see Ref.\cite{BRPBS}. 
I make a better case (I hope) for continuing probing {\bf CP} violation in $\tau$ decays.

Now I discuss {\bf CP} asymmetries in $\tau$ decays with some details. To be realistic (from a theorist's view)  I talk about 
$\tau^- \to \nu K^- [\pi^0/\pi^+\pi^-/\eta]$ and $\tau^- \to \nu \bar K^0 [\pi^-/\pi^-\pi^0/\pi^-\eta ]$ with somewhat similar branching ratios, 
see the PDG2020 data. 
However, the SM describes quite differently the `landscape' for {\bf CP} asymmetries.

\vspace{3mm}

Class (I) 
\bea
{\rm BR} (\tau^- \to \nu K^- \pi^0) & = &  (4.33 \pm 0.15 ) \cdot 10^{-3} 
\label{K-PI0}
\\
{\rm BR}  (\tau^- \to \nu K^- \pi^+ \pi^-) & = & (3.45 \pm 0.07 ) \cdot 10^{-3}  
\label{K-PI+PI-}
\\
{\rm BR}  (\tau^- \to \nu K^- \eta) & = &  (1.55 \pm 0.08 ) \cdot 10^{-4} 
\label{K-ETA} 
\eea

The SM gives zero {\bf CP} violation in these channels.   

\vspace{3mm}

Class (II)
\bea
{\rm BR} (\tau^- \to \nu \bar K^0 \pi^-) & = &  (8.38 \pm 0.14 ) \cdot 10^{-3} 
\label{K0PI}
\\
{\rm BR}  (\tau^- \to \nu \bar K^0 \pi^- \pi^0) & = & (3.82 \pm 0.13 ) \cdot 10^{-3}  
\label{K0PI-PI0}
\\
{\rm BR}  (\tau^- \to \nu \bar K^0 \pi^- \eta) & = &  (0.94 \pm 0.15 ) \cdot 10^{-4} 
\label{KPIETA} 
\eea

The situation is more `complex' with neutral kaon about {\bf CP} asymmetries: 
\bea
A_{\bf CP} (\tau^- \to \nu K^0_{S} \pi^-)|_{\rm PDG2020} & = & + \; (0.36 \pm 0.25 ) \% 
\label{PDG2020}
\\ 
A_{\bf CP} (\tau^- \to \nu K^0_S \pi^-)|_{\rm SM} & = & - \; (0.36 \pm 0.01) \%     \; . 
%\; \; \; \; \; \cite{BSGN}
\label{SMPRED}
\eea
I give reference to two articles about {\bf CP} asymmetry in $\tau^- \to \nu K_S\pi^-$ in the past  
\cite{BSGN} 
\footnote{I am not a super-fan of thinking about a connection between $\tau^-$ vs. $D^+$ transitions.}.  

Which lesson  one can learn here? 
\begin{itemize}
\item
So far our community has not established impact of ND on $A_{\bf CP}(\tau^- \to \nu K^0_S \pi^-)$; i.e., the 
SM prediction is consistent with the PDG2020 data within 3$\sigma$, see Eq.(\ref{SMPRED}) vs. Eq.(\ref{PDG2020}). 

\item
The situations are very different between Class (I), where the SM predicts zero {\bf CP} violation, and Class (II), where it predicts 
$A_{\bf CP} (\tau^- \to \nu K^0_S \pi^-)|_{\rm SM} =  - \; (0.36 \pm 0.01) \cdot 10^{-3}$. Actually this SM asymmetry is subtle: it is based on 
{\em indirect} {\bf CP} violation in neutral $K \to \pi^+\pi^-$, although in general $\tau$ decays could show {\em direct} {\bf CP} asymmetries. 

\item
For phenomenalists the goals are to convince experiments not only to probe {\bf CP} asymmetry in $\tau^- \to \nu K_S \pi^-$ with accuracy, but also 
with a broader program: 
\\
$\tau^- \to \nu K^0_S [\pi^- \pi^0/ \pi^- \eta]$ and $\tau^- \to \nu K^-[\pi^0/\pi^+\pi^-/\eta]$.

%When I was working on my contribution to the TAU2021 workshop, I had realized that my 2021 book has giving less than one page for 
%{\bf CP} asymmetry, namely for $\tau^- \to \nu K_S \pi^-$, while around three pages in my 2009 book, see Ref.\cite{BRPBS}. 
%I make a better case (I hope) for continuing probing {\bf CP} violation in $\tau$ decays.    

\item
A small  comment: our community can probe {\bf CP} asymmetry in Cabibbo favored transitions like $\tau^- \to\nu K^- K^0$ 
with the opposite sign in the SM; furthermore it could show impact of ND. Of course, one can compare the branching ratios: 
\bea
{\rm BR}(\tau^- \to \nu \pi^- \pi^0)  & = & (25.49 \pm 0.09) \; \% 
\\
{\rm BR}(\tau^- \to \nu K^- K^0)  & = &  (1.486 \pm 0.034 ) \cdot 10^{-3}  \; .
\eea
At the TAU2012 workshop I had mostly focussed on $\tau^- \to \nu K\pi$'s, but I had mentioned FS with only $\pi$'s. \cite{BIGITAU2012}. 

The SM predicts zero {\bf CP} asymmetries in $\tau^- \to \nu K^-[\pi^0/ \pi^+\pi^-]$.  
The situations are somewhat similar for doubly Cabibbo suppressed $D^- \to K^- \pi^0/K^-\pi^+\pi^-$.  
Thus there are `hunting areas' for impact of ND on {\bf CP} asymmetries in $\tau$ \& $D^-$ decays.

\item
Kiers in his TAU2012 talk had discussed  both FS of $K\pi$'s and $\pi$'s \cite{KIERS}.   
 
For mostly practical reasons they had discussed  {\bf CP} asymmetries first in $\tau^- \to \nu \pi$'s; it had suggested that model  
with  charged Higgs fields could hardly have impact on $\tau^- \to \nu \pi^-\pi^0$, but on $\tau^- \to \nu [3 \pi/4\pi]$ \cite{DATTA2007}.

\item\
Again as a phenomenalist (in the world of quarks \& leptons): I `paint' the `landscape': 
$\tau^- \to \nu \; s ... \bar u \Rightarrow \nu K^- \pi^+\pi^-/\nu \bar K^0  \pi^-\pi^0$ etc. Somewhat similar situation with Cabibbo favored transition  
$\tau^- \to \nu \; d ... \bar u \Rightarrow \nu  K^- K^0 $.

Due to $K^0 - \bar K^0$ the SM predicts
\beq
A_{\bf CP} (\tau^- \to \nu K_S \pi^- ) = + \; (0.36 \pm 0.01) \% = - A_{\bf CP} (\tau^- \to \nu K_S K^- )
\label{CPVK-K_S}
\eeq

\item
"Leptoquarks" ({\em LQ}) have been become as a `fashion', in particular for the theoretical literature about their indirect impact.  One can look at 
$\tau^- \to s \; "LQ" \; \nu \; \bar u \Rightarrow  \nu \bar K^0 \pi^-/\nu K^-\pi^+\pi^- $ or even $\tau^- \to d \; "LQ" \; \nu  \; \bar u \Rightarrow \nu K^- K^0$ 
that could violate Eq.(\ref{CPVK-K_S}).

\end{itemize}
The situations about {\bf CP} asymmetries in $\tau$ transitions are very different about {\bf CP} violations in the decays of beauty, charm \& 
strange mesons at least, where SM had predicted non-zero values.  
Here I am not talking about numbers or models or even classes of models. As a phenomenalist one would say in one channel that {\bf CP} 
asymmetry has been  established the impact of ND. However, one has to be realistic in our world with experimental data; one has to 
find possible patterns; thus I use again the word of `paintings', 
see {\bf Figure \ref{MONTSTMICHEL}}:
%%%%%%%%%
\begin{figure}[!h]
\begin{center}
\includegraphics[width=6cm]{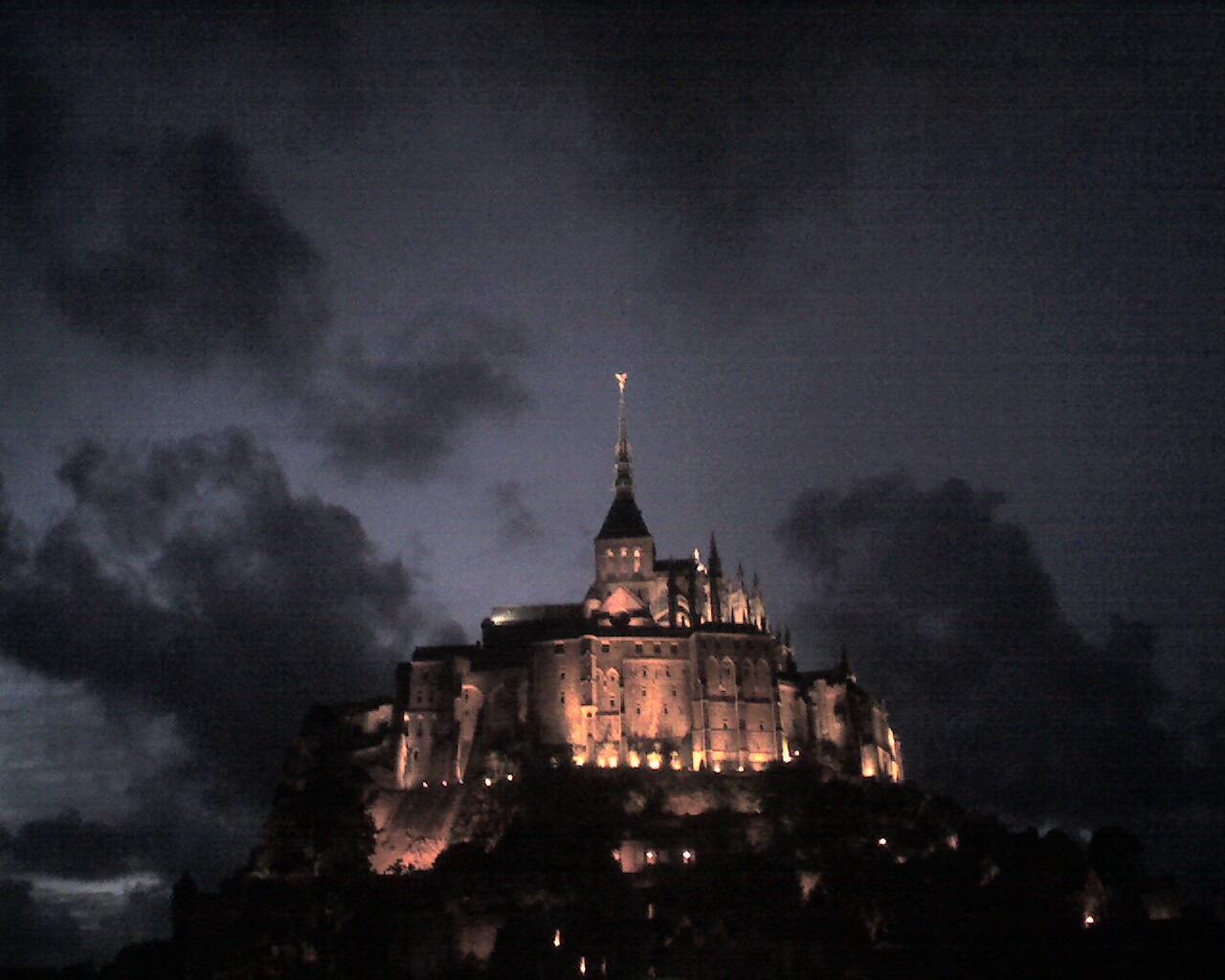}
\end{center}
%\vspace{-9.0cm}
%\vspace{-0.5cm}
\caption{Mont Saint-Michel is an island in Normandy (France) [picture taken by IIB].}
\label{MONTSTMICHEL}
\end{figure}
%%%%%%%%%%%

My summarizing of the 2020 data:
\begin{itemize}
\item
PDG2020 data are consistent with SM predictions including {\bf CP} violation.

\item
However, the situation is `thin' about {\bf CP} violation in $\tau$ decays: PDG2020 gives only for 
$ A_{\bf CP} (\tau^- \to \nu K^0_{S} \pi^-)|_{\rm PDG2020}  =  + \; (0.36 \pm 0.25 ) \% $. 

It is important to probe {\bf CP} asymmetry in different FS Eq.(\ref{K-PI0}) - Eq.(\ref{KPIETA})   (as listed above). 
I think it is an excellent candidate for the {\em bronze} medal.

\end{itemize}
At the TAU2021 WS `we' had been told that the Belle II collaboration is `thinking' about {\bf CP} asymmetries in $e^+ e^- \to \tau^+ \tau ^-$, which is good. 
What are the limits in different channels?

One can find a very nice 2013 review by Pich with the title "Precision Tau Physics" with 50 pages (except the list of References) 
\footnote{It gives a list of 601 references which helps readers to understand the underlying $\tau$ dynamics.} \cite{PICH2013}. 
When one looks at the `landscape' of $\tau$ dynamics in a review, the `Sect.10 CP violation'  could be discussed in less than one page; 
however, 16 references were given there \footnote{It is discussed also in Sect.9 `Electromagnetic and Weak Dipole Moments'}.

%%%%%%%%%%%%%%%%%%%%% 
\section{Probing $\tau$ transitions with polarized electron beams} 
\label{POLARIZED}
%%%%%%%%%%%%%%%%%%%%%%

I list articles about {\bf CP} violation in $e^+ e^- \to \tau^+ \tau^-$ with polarized $e^-$ beams,  
see Rets.\cite{TIM,BIGITAU2012,KIERS,PICH2013} or mentioned in Ref.\cite{BSPolBeam}. 
It was pointed out by Bernabeu {\em et al.} in Ref.\cite{BERNABEU2007}  to focus on leptonic EDMs for good reasons, and I have to agree. 
Of course, it is a true challenge to produce $e^+ e^-_{\rm polar.} \to \tau^+\tau^-$. 
I tell about a `broader' (but maybe not deeper) `landscape', namely {\bf CP} asymmetries in $\tau$ decays.  

One can notice that the articles came from theorists. However, the situation has changed. 
We were able to listen very nice talks  by X. Zhou  about a possible Super Tau-Charm Factory in China \cite{CHINASuper} 
and by D. Epifanov about a Russia one  \cite{RUSSIASuper,DENIS2020}.  
One can read some details in their contributions.

It would have rich programs in general; furthermore, it opens new `roads' 
about fundamental dynamics in $\tau$ EDMs and {\bf CP} asymmetries. I only somewhat disagree: `we' should not focus only  on {\bf CP} asymmetries 
on $\tau^- \to \nu (K\pi)$; we have to probe also {\bf CP} asymmetries in $\tau^- \to \nu K[\pi\pi/\eta/\pi\eta]$, as I said above including 
$e^+e^-_{\rm polar.}\to \tau^+\tau^-$. Of course, the first goal is to find {\bf CP} asymmetry in $e^+e^- \to \tau^+\tau^-$. Later one can use 
a $e^-_{\rm polar.}$ beam: it is not about bragging rights about an experiment; one could find a deeper information about the underlying dynamics.

%%%%%%%%%%%%%%%%%%%%%%%
\section{Lessons for the future} 
\label{MOTTO}
%%%%%%%%%%%%%

I give a short comment about {\bf CP} asymmetries in the decays of {\em strange baryons} like for $\Lambda \to p \pi^-$:
I am a co-auther in a 2018 article \cite{CPVSTRBARY}. 

The SM situations of {\bf CP} violation are very different between 
$\tau^-$ and $\Lambda$ (in principle; in real data one has to worry about uncertainties in the measured values): 
\beq
A_{\bf CP}(\tau^- \to K^- [\pi'{\rm s}]) = 0  \; \; \; \;  {\rm vs.} \; \; \; \;  A_{\bf CP}(\Lambda \to p \pi^-) \neq 0 \; ; 
\eeq
so far we have no real SM prediction for $ A_{\bf CP}(\Lambda \to p \pi^-)$. 
It was listed as an important goal for a future Super Tau-Charm Factory  \cite{CHINASuper}. 

Mostly our literature about leptonic EDM's is for electrons \& muons; however it is important to probe $\tau$ EDMs in a different `landscape', 
as discussed in the 2021 Ref.\cite{BERN} with the title "Electric dipole moment of the tau lepton revisited".  

On the other hand, I focus on {\bf CP} asymmetries in $\tau$ decays. The SM predicts zero {\bf CP} asymmetries in $\tau^-$
transitions (except {\bf CP} violation in the decays of neutral kaons). As I had said above, {\bf CP} asymmetries in 
semi-hadronic  $\tau^-$ decays are good candidate for the bronze medal. 
The hadronic FS are described including resonances: some are narrow, while other are broad. They are `complex'
in the world of quarks, see the `painting' of $s ... \bar u$   or $d ... \bar u$ leading to two \& three mesons (or even four ones). 

I list three classes of $\tau^-$ decays about {\bf CP} asymmetries: 
(a) $\tau^- \to \nu K^- [ \pi^0/\pi^+\pi^-/\eta]$; \\
(b) $\tau^- \to \nu [\pi^-\pi^0/\pi^-\pi^+\pi^-/\pi^- \eta] $; 
(c) $\tau^- \to \nu K_S [ \pi^-/\pi^-\pi^0/\pi^-\eta]$ \footnote{One could include $\tau^- \to \nu K^0K^-$.}. The SM predicts zero {\bf CP} asymmetries 
in classes (a) \& (b), while non-zero value due to {\bf CP} violation in the measured $\bar K^0 - K^0$ oscillations.  
\begin{itemize}
\item
{\bf CP} violation is an excellent tool for finding ND due to an amplitude in $\tau$ rates.

\item
The FS of semi-hadronic $\tau$ decays have one, two, three ... mesons. `We' have decent chance to find ND in {\bf CP} 
asymmetries, in particular beyond $\tau^- \to \nu h_1h_2$.  For example: $\tau^- \to \nu K^- \pi^+\pi^-$.  

The situations about {\bf CP} violation are very different between the $\tau$ lepton vs. the strange, beauty \& charm mesons 
already on the qualitative level.

\item 
One can find models for {\bf CP} asymmetries for $\tau$ transitions (like with charged Higgs or leptoquarks fields) that can produce {\bf CP}
asymmetries, but my main points are: future experiments should probe 
{\bf CP} violation  in $\tau^- \to \nu (h_1h_2h_3)$ decays with $h_i$ mesons 
\footnote{As usual, the hard work has to be done from experimenters.}.

\end{itemize}
Once we get new limits or even better to get non-zero values, we can discuss what we have learnt about underlying dynamics 
like chiral symmetry. Thus it is crucial to use collaboration with members of HEP vs. Hadrodynamics 
\footnote{Hadrodynamics is a much better choice of words than MED. } from different `cultures'.

%%%%%%%%%%%%%%%%%%%%%%%
\section{Epilogue} 
%%%%%%%%%%%%%

One can go to old history: "Gods = Symmetries speak in Riddles." 
Or: "On seeing a missile shot by a catapult which had been brought them for the first time, a king from Sparta in the 4th century B.C. cried out: 
`By Heracles, this is the end of man's valor'." Can a theorist see an analogy with computers? 

%%%%%%%%%%
\begin{figure}[h!]
\begin{center}
\includegraphics[width=4cm]{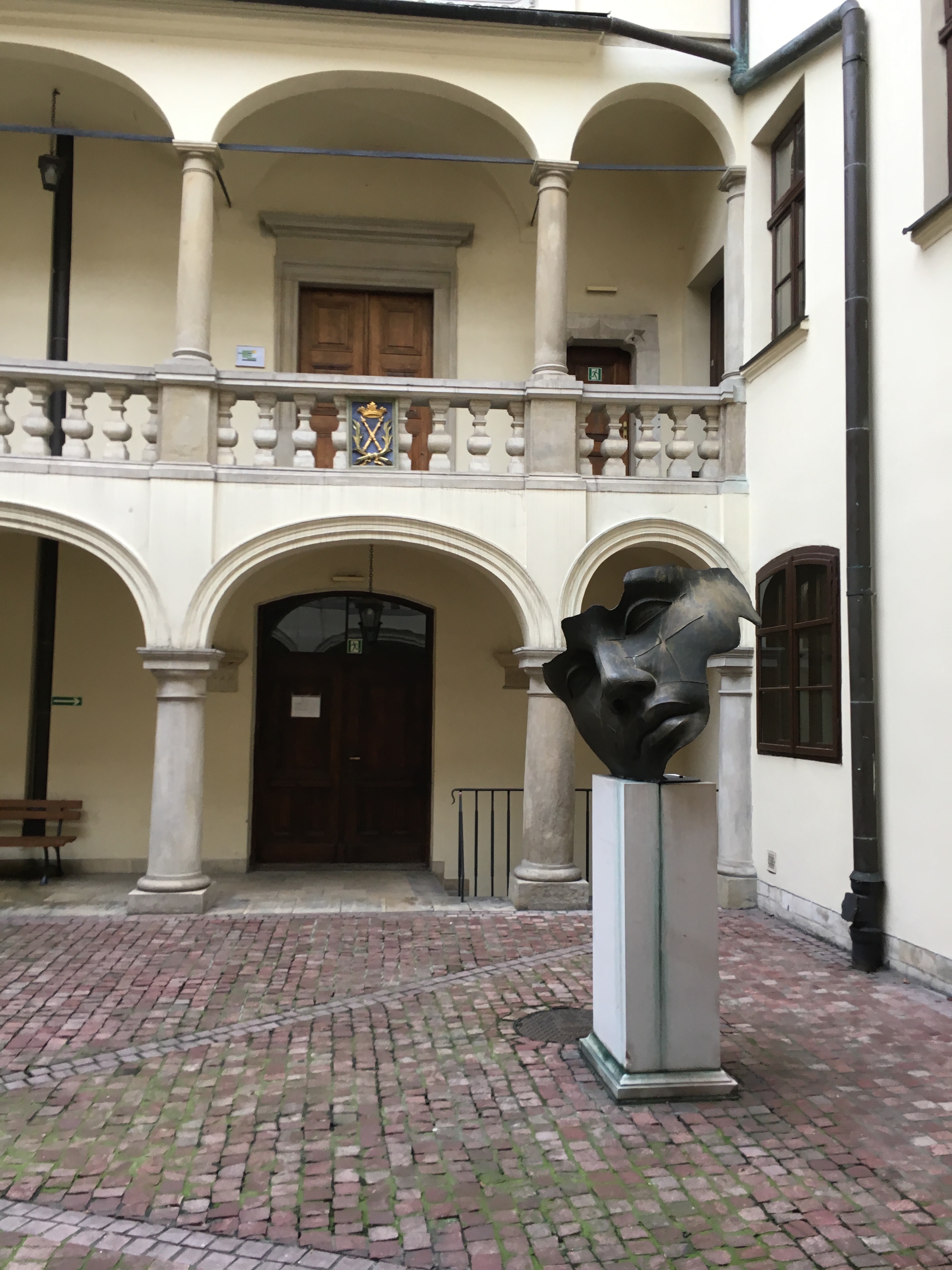}
\end{center}
 \vspace{0cm}
\caption{Renaissance architecture \& modern sculpture (picture taken by IIB)}
\label{2arts}
\end{figure}
%%%%%%%%%%%%%%%%%

\noindent
I want to show an analogy of {\bf CP} asymmetries in $\tau$ transitions with `art', see the {\bf Figure \ref{2arts}}. It is wonderful connection of Renaissance architecture and modern sculpture. It is inside a building  just south of the Main Market Square of the city Cracow (Poland). It is not easy to find it; 
thus I am proud of this picture.

Our understanding of fundamental dynamics is based on symmetries in many ways --
although often that is not  obvious. 
We can learn from paintings (or other arts). Some artists can have `visions', not just describe the landscape: when one talks about 
{\em triangles}, {\em quadrangles}
etc. in the different flavor dynamics of quarks and leptons,  the paintings of the Russian artist Kandinsky come to mind: 
one can look at page 87 in Ref.\cite{BRPJuly2021} 
from Kandinsky's 1923 paintings before quantum mechanics, see {\bf Figure \ref{RUSSIANARTIST}}:
%see {\bf Fig.\ref{fig:RUSSIANTRIANGLES}}
%%%%%%%%%%
\begin{figure}[h!]
\begin{center}
\includegraphics[width=6cm]{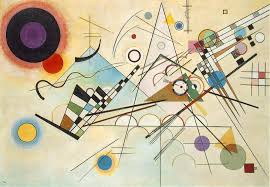}
\end{center}
\vspace{0cm}
\caption{ "Composition VIII"  (painted by V., Kandinsky in 1923)}
\label{RUSSIANARTIST}
\end{figure}
%%%%%%%%%%%

\noindent
Another example: this time from a Spanish painter, namely Salvador Dali. At my talk at the TAU2021 workshop I had shown his 1931 painting  
with the title "Persistence of Memory". Here I show his 1937 painting, see  {\bf Figure \ref{Dali}}:
%%%%%%%%%%
\begin{figure}[h!]
\begin{center}
\includegraphics[width=6cm]{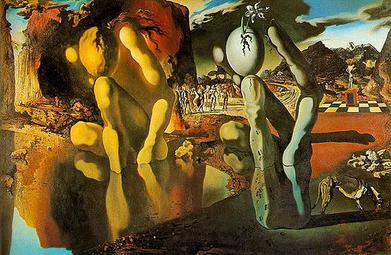}
\end{center}
 \vspace{0cm}
\caption{ "Metamorphosis of Narcissus"  (painted by S. Dali in 1937)}
\label{Dali}
\end{figure}
%%%%%%%%%%%%%

{\bf My final comment here:} As I had said above theorists can produce models of ND that could be found  
{\bf CP} asymmetries in $\tau^{\pm}$ transitions for the next decade.  So far, I am not convinced by these models. 
I am happy  to hear about {\bf CP} asymmetries in $\tau$ decays by the LHCb and the Super-Tau-Charm Factory. However, 
`our'   community has to {\em beyond} $\tau^- \to \nu (K\pi)^-$ -- it is very important. 
There is an analogy between `real art' and `fundamental physics'.

%%%%%%%%%%
%\begin{figure}[h!]
%\begin{center}
%\includegraphics[width=6cm]{dalitime.jpg}
%\end{center}
 %\vspace{0cm}
 %\caption{ "Time \& Change" (painted by Dali)}
%\label{dalitime}
%\end{figure}see Fi
%%%%%%%%%%%%%%%%

\section*{Acknowledgements}
I appreciate discussions I had with X. Zhou \& F. Noel and in particular with D. Epifanov, see in Refs.\cite{RUSSIASuper,DENIS2020}.   
This work was supported by the NSF PHY-1820860.

%%%%%%%%%%%%%%
%\section{References}
%%%%%%%%%%%%%%

%$\nolinenumbers

\end{document}